\documentclass[12pt]{article}

\usepackage{amssymb}

\begin{document}

\begin{titlepage}

\begin{flushright}
\end{flushright}
\vskip 2.5cm

\begin{center}
{\Large \bf Top Hadrons in Lorentz-Violating Field Theory}
\end{center}

\vspace{1ex}

\begin{center}
{\large Brett Altschul\footnote{{\tt altschul@mailbox.sc.edu}}}

\vspace{5mm}
{\sl Department of Physics and Astronomy} \\
{\sl University of South Carolina} \\
{\sl Columbia, SC 29208} \\
\end{center}

\vspace{2.5ex}

\medskip

\centerline {\bf Abstract}

\bigskip

If there is Lorentz symmetry violation in the $t$ quark sector of the standard model, changes to particles'
dispersion relations might allow for the existence of stable top-flavored hadrons. Observations of the survival
of high-energy $\gamma$-rays over astrophysical distances
can be used to place one-sided constraints on certain linear combinations of Lorentz violation coefficients
in the $t$ sector at the $\sim 10^{-4}$ level of precision.

\bigskip

\end{titlepage}

\newpage

\section{Introduction}

Since the introduction of the special theory of relativity, there has always been interest in understanding
whether the Lorentz symmetry of special relativity is exact, or whether it is merely an excellent approximation.
Approximate symmetries have turned out to be an extremely important topic in
particle physics. Thanks to advances in effective field theory over the last quarter century, there has
been an explosion of interest in the quantitative problem of evaluating how well Lorentz invariance has been
confirmed.
The fundamental physics that we currently understand is based on two very different theories,
the standard model, describing particle physics, and general relativity, which describes gravitation.
Understanding how these two theories can be reconciled into a theory of quantum gravity
is probably the greatest remaining challenge in fundamental physics. However, despite their puzzling
differences, the two basic theories actually have a number of important features in common. These
include a number of spacetime symmetries; both the standard model and general relativity are
invariant under rotations, Lorentz boosts, and CPT.
Experimental searches have not turned up any compelling evidence of Lorentz violation; however, it
remains a topic of theoretical and experimental interest. If any deviations from Lorentz invariance were
conclusively demonstrated, that would be a discovery of tremendous significance and would open a new window
into our understanding of the laws of physics at the most elemental level.

Using the machinery of effective theory, it is possible to describe deviations from these symmetries, whether
involving standard model quanta or gravitational effects.
The general effective field theory that can be used to describe such effects
is called the standard model extension (SME). The full SME contains an infinite hierarchy of Lorentz-violating
operators. However, the action for the minimal SME---which contains only those operators constructed out of known standard
model fields that are gauge symmetric,
translation invariant, and power-counting renormalizable---contains a finite number of coupling parameters.
The SME operators resemble the operators that appear in the usual standard model action, except that
they may break Lorentz symmetry when they have residual tensor indices. The Lagrange density may be written in a form
in which the residual indices on the operators are contracted with tensor-valued coefficients, so that the coefficients
indicate the presence of preferred vector and tensor backgrounds~\cite{ref-kost1,ref-kost2}.
If the origin of the Lorentz violation is a form of spontaneous symmetry breaking, then the preferred background
tensors are set by the vacuum expectation values of tensor-valued dynamical fields.

The minimal SME is well suited for comparing the results of experimental Lorentz tests. Experiments in many
different areas of physics have been used to place bounds on the minimal SME
coefficients. Information about the best current bounds are collected and summarized in~\cite{ref-tables}.
It happens that many of the strongest bounds on various SME coefficients come from astronomy. These bounds typically
take advantage of either the exceeding long travel distances or very high energies that are available to
certain extraterrestrial quanta. Observations of photons from astrophysical sources have thus been used to place many of
the best constraints on Lorentz violation.

By examining the photons emitted by very distant sources, we may learn a
great deal about the dispersion relation of the photons themselves, to see whether it has the expected
Lorentz-invariant form $E_{\gamma}=|\vec{p}\,|$. However, observing that a photon was emitted and that it made its way
to our detectors can also tell us quite a bit about the behavior of the other particles that might interact with such
a photon. The observation of TeV-scale $\gamma$-rays reveals information about the processes that might have produced
them---typically either inverse Compton scattering, $e^{-}+\gamma\rightarrow e^{-}+\gamma$, involving the upscattering
of a low-energy photon by an exceedingly  energetic electron, or neutral pion decay, $\pi^{0}\rightarrow\gamma+\gamma$.
The fact that such photons are produced at all tells us things about energy-momentum relations of
all the particles that are involved in the reaction. The experimental observation of the spectra of astrophysical
$\gamma$-ray sources can therefore be used to place bounds on Lorentz violation in the lepton and hadron sectors
of the SME~\cite{ref-jacobson1,ref-altschul6,ref-altschul15,ref-altschul16}.

However, even when the production methods of these $\gamma$-rays are unclear, they can still tell us interesting
things about Lorentz violation in other sectors, merely based on the fact that the photons live long enough to reach us.
For the massless photon, any kind of decay process that occurs in vacuum and
involves one or more massive daughter particles is forbidden by energy-momentum conservation, if all the
quanta have their usual Lorentz-invariant dispersion relations. A decay such as $\gamma\rightarrow e^{+}+e^{-}$ or
$\gamma\rightarrow t+\bar{t}$ (with a top-flavored quark-antiquark pair),
or an emission process such as $\gamma\rightarrow\gamma+\pi^{0}$,
cannot occur without another constituent (such as a nearby heavy nucleus)
to take up the photon's excess momentum. However, if the energy-momentum relations are Lorentz violating---so
that, at large momenta, the energies of the daughter particles increase more slowly as a function of $p$ than
the energy of the parent photon $E_{\gamma}(\vec{p}\,)$---then the process may be permitted above some
threshold. (Decay of a photon into other strictly massless quanta---that is, photon splitting,
$\gamma\rightarrow N\gamma$---is separately forbidden, not by energy-momentum conservation, but by
vanishing of the relevant on-shell matrix element~\cite{ref-schwinger1}. Yet for that kind of process also,
Lorentz-violating radiative corrections may again make the process possible, even when then photon sector itself is not
modified~\cite{ref-kost5}.
However, since the process is always precisely at threshold, there are phase space issues, and the process would
manifest itself not as an irreversible decay, but via oscillation phenomena, such as
$\gamma\rightarrow2\gamma\rightarrow\gamma\rightarrow\cdots$.)

In this paper, we shall expand the use of observations of the survival of ultra-high-energy propagating photons to
place constraints on Lorentz violation in the $t$ sector. Section~\ref{sec-SME} describes the structure and
one-particle kinematics of the minimal SME operators of interest. The kinematic details of the single-photon
$t$-$\bar{t}$ production process,
and the bounds on SME parameters that result from the observed absence of this process are discussed in
section~\ref{sec-kinematics}. The $t$ sector is more complicated than others, because
$t$ quarks do not normally exist as constituents of asymptotic hadron states; a $t$ produced in a reaction
will ordinarily decay weakly before
hadronization occurs. This will make the analysis for this sector somewhat more intricate than that required in
other sectors of the SME. Our conclusions, and the outlook for further improvements, are discussed in
secton~\ref{sec-concl}.

\section{Lorentz Violation in the Minimal SME}
\label{sec-SME}

The Lagrange density for the minimal SME contains operators of mass dimensions 3 and 4, with their
outstanding indices contracted with those of tensor-valued coefficients. For the QED sector of the minimal SME,
with a single fermion species, the Lagrange density is
\begin{equation}
\label{eq-L}
{\cal L}=-\frac{1}{4}F^{\mu\nu}F_{\mu\nu}
-\frac{1}{4}k_{F}^{\mu\nu\rho\sigma}F_{\mu\nu}F_{\rho\sigma}
+\frac{1}{2}k_{AF}^{\mu}\epsilon_{\mu\nu\rho\sigma}F^{\nu\rho}
A^{\sigma}+\bar{\psi}(i\Gamma^{\mu}D_{\mu}-M)\psi,
\end{equation}
where the SME coefficients for the fermion are collected in 
\begin{eqnarray}
M & = & m+a^{\mu}\gamma_{\mu}+b^{\mu}\gamma_{5}\gamma_{\mu}+\frac{1}{2}H^{\mu\nu}\sigma_{\mu\nu}+im_{5}
\gamma_{5} \\
\Gamma^{\mu} &= & \gamma^{\mu}+c^{\nu\mu}\gamma_{\nu}+d^{\nu\mu}\gamma_{5}
\gamma_{\mu}+e^{\mu}+if^{\mu}\gamma_{5}+\frac{1}{2}g^{\lambda\nu\mu}
\sigma_{\lambda\nu}.
\end{eqnarray}
The gauge-covariant derivate $D_{\mu}=\partial_{\mu}+iqA_{\mu}$ contains the charge of the fermion species,
with $q=-\frac{2}{3}e=\frac{2}{3}|e|$ for the $t$ quark.

At ultrarelativistic energies, the dimension-4 coefficients ($k_{F}$ for photons and those comprising $\Gamma$
for the fermion field) dominate the
dispersion relations for the quanta, so we may neglect all the dimension-3 operators except the usual
$M\approx m$. Moreover, while the $e^{\mu}$, $f^{\mu}$, and
$g^{\lambda\nu\mu}$ coefficients are acceptable in a pure QED theory, they are not
consistent with the $SU(2)_{L}$ gauge invariance of the standard model. They can only arise as vacuum expectation
values of higher-dimensional operators involving the Higgs field and thus actually excluded from the truly minimal SME.
This leaves the fermion-sector Lorentz violation at ultrarelativistic energies predominantly controlled by
$c^{\nu\mu}\gamma_{\nu}+d^{\nu\mu}\gamma_{5}\gamma_{\mu}$. For the electromagnetic sector, the dimension-4
$k_{F}^{\mu\nu\rho\sigma}$
tensor (which, in the most general case, may be taken to have the same symmetries as a Riemann curvature tensor
and a vanishing double trace)
may be split into two-parts (with Weyl-like and Ricci-like structures, respectively).
The ten-components of the Weyl-like part
(as well as the four possible dimension-3 $k_{AF}^{\mu}$ terms) give rise to photon birefringence in vacuum, which has not
been seen, even for polarized sources at cosmological distances~\cite{ref-mewes5,ref-mewes6}. The bounds on certain
linear combinations of these dimensionless coefficients are down at the $10^{-38}$ level. We shall therefore neglect all
these birefringent terms. The remaining Ricci-like terms have the structure
\begin{equation}
k_{F}^{\mu\nu\rho\sigma}=\frac{1}{2}\left(g^{\mu\rho}\tilde{k}_{F}^{\nu\sigma}
-g^{\mu\sigma}\tilde{k}_{F}^{\nu\rho}
-g^{\nu\rho}\tilde{k}_{F}^{\mu\sigma}
+g^{\nu\sigma}\tilde{k}_{F}^{\mu\rho}\right),
\end{equation}
where $\tilde{k}^{\mu\nu}=k_{F\alpha}\,^{\mu\alpha\nu}$.
The bounds on these $\tilde{k}_{F}^{\mu\nu}$
coefficients are weaker, although for some terms, the bounds are still quoted at the
at the quite respectable $10^{-22}$ level~\cite{ref-mewes7}, based on a Michelson-Morely analysis of
Laser Interferometer Gravitational-Wave Observatory (LIGO) data (although it should be noted these bounds are actually
necessarily on differences between the electromagnetic $\tilde{k}_{F}^{\mu\nu}$ coefficients and some set of aggregate
coefficients for everyday matter).

With flavor-changing SME coefficients neglected, there are separate sets of background tensors for each fermion
species. However, because of the $SU(2)_{L}$ gauge invariance, which connects fermions with different charges,
not all the coefficients for different flavors are truly independent. Between the $t$ and $b$ quarks, there are really only
three independent $c^{\mu\nu}$ and $d^{\mu\nu}$ tensors. The gauge invariance ensures that
the chiral SME coefficint $c_{L}^{\mu\nu}=c^{\mu\mu}+d^{\mu\nu}$ is the
same for the left-handed $t$ and $b$ quarks that form an $SU(2)_{L}$ doublet. However, the 
$c_{R}^{\mu\nu}=c^{\mu\mu}-d^{\mu\nu}$ for the two species are, generally speaking, unrelated.

The tensor $\frac{1}{2}\tilde{k}_{F}^{\mu\nu}$ is the photonic analogue of the $c^{\mu\nu}$ for fermions. Each of them
represents a spin-independent modification of the ultrarelativistic dispersion relation.
For the fermions, the modified dispersion relation is most straightforwardly
expressed in terms of the maximum achievable velocity (MAV) of a particle (which can depend on
the direction of its motion and its spin). The MAV along the direction $\hat{v}$ is
$1+\delta(\hat{v})$, where~\cite{ref-altschul4}
\begin{equation}
\label{eq-delta}
\delta(\hat{v})=-c_{00}-c_{(0j)}\hat{v}_{j}-c_{jk}\hat{v}_{j}\hat{v}_{k}+sd_{00}+sd_{(0j)}
\hat{v}_{j}+sd_{jk}\hat{v}_{j}\hat{v}_{k},
\end{equation}
and where $s$ is the product of the particle helicity and fermion number (e.g. $+1$ for a quark, $-1$ for
an antiquark). The parentheses, as in $c_{(0j)}$, indicate a symmetrized sum 
$c_{0j}+c_{j0}$. Note that $c^{\mu\nu}$ and $d^{\mu\nu}$ appear in $\delta(\hat{v})$ only via contractions
$c^{\mu\nu}v_{\mu}v_{\nu}$ and $d^{\mu\nu}v_{\mu}v_{\nu}$, where $v^{\mu}=(1,\hat{v})$.
The dispersion relation for a relativistic
quantum is
\begin{equation}
E=\sqrt{m^{2}+p^{2}[1+2\delta(\hat{p})]},
\end{equation}
up to corrections of ${\cal O}\left(m^{4}/E^{4}\right)$. For one-dimensional motion restricted to the
$\hat{p}$-direction, this just looks like ordinary special relativity, but with a modified top speed
$1+\delta(\hat{v})$. The actual velocity at these ultrarelativistic energies is
\begin{equation}
\label{eq-v}
\vec{v}=\left[1+\delta(\hat{p})-\frac{m^{2}}{2p^{2}}\right]\hat{p},
\end{equation}
which is simply the usual ultrarelativistic
velocity plus the Lorentz-violating correction to the MAV,
$\delta(\hat{p}\,)\hat{p}$.

An electromagnetic decay such as $\gamma\rightarrow t+\bar{t}$ would occur extremely rapidly if it were allowed.
The survival of GeV-scale laboratory photons over everyday distances has been used to place a bound on certain
SME coefficients~\cite{ref-hohensee1}. So
if a photon, with a certain energy and traveling in a certain direction, manages to reach us from an astronomical source,
that is essentially certain evidence that there are no photon decay products that are kinematically permitted
for that particular combination of energy and direction.
It is well established that photons with TeV energies can survive over astrophysical distances, and this has
been used to place bounds on the differences between certain Lorentz violation coefficients for photons and
the kinematically
equivalent coefficients for leptons or hadrons~\cite{ref-stecker,ref-altschul14}. The absence of the process
$\gamma\rightarrow X^{+}+X^{-}$ in a direction $\hat{v}$ places restrictions on the linear combination
$\left(c_{X}^{\mu\nu}-\frac{1}{2}\tilde{k}_{F}^{\mu\nu}\right)\!v_{\mu}v_{\nu}$ of SME parameters. The strength of the
bound depends on the mass $m_{X}$ of the charged daughter particles, with a characteristic
${\cal O}(m_{X}^{2}/E_{\gamma}^{2})$ dependence. The strongest individual bounds
come from the photons with the highest observed energies, but $\gamma$-rays coming from a variety of different directions
are required to get the best overall constraints on all the components of the traceless, symmetric Lorentz tensor
$c_{X}^{\mu\nu}$.

Lorentz violation for photons, and for the first-generation fermions that are the primary fermionic constituents of
everyday matter, is generally fairly well constrained. Bounds for second- and, especially, third-generation
fermions are much less comprehensive. It can be quite challenging to do high-precision experiments with
short-lived particles.
Lorentz violation for the top sector is currently rather poorly constrained, with all the existing bounds
in the literature being derived in one of two fashions. There are direct measurements of the $t$-$\bar{t}$
production cross section~\cite{ref-abazov}, with sensitivities to the coefficients of dimension-4 SME
operators at the $10^{-1}$--$10^{-2}$ level. The fact that $t$ quarks do not exist as external states is not an
issue for these measurements, because the SME coefficients can affect the physical cross sections not primarily through
the kinematics but also directly via
the dynamical matrix element for the $t$-$\bar{t}$ production
process~\cite{ref-berger4}. The other bound arises from radiative corrections. The $t$ quark field, like all other
charged fermion fields, can appear via a virtual particle-antiparticle pair in the vacuum polarization diagram
of the photon self-energy. This will transfer any Lorentz violation in the charged matter sector to the photon
sector, including the radiative generation of dimension-6 operators in the electromagnetic Lagrange density.
An analysis of these higher-dimensional radiative corrections---generalizing earlier perturbative demonstrations
of the one-loop renormalizability of the quantum electrodynamics (QED) sector of the
minimal SME~\cite{ref-kost3}---placed a bound on a dimension-4 SME operator
in the $t$ sector at the $10^{-7}$ level, based on the relative arrival times of TeV $\gamma$-rays
from an active galactic nucleus~\cite{ref-satunin}. 
However, the calculations leading to this bound relied an an assumption
of spatial isotropy. Moreover, since radiative corrections from multiple species are
involved, there could potentially be cancelations between the effects of different types of virtual particles.
The bound is thus best characterized as a constraint on
a particular SME coefficient $\left|c_{t}^{00}\right|$, under the rather
restrictive assumption that this is the only nonzero SME parameter.

\section{Photon Decay into Top-Flavored Hadrons}
\label{sec-kinematics}

Normally, when a $t$-$\bar{t}$ pair is produced at an accelerator, the large mass $m_{t}$ ensures that
the quark and antiquark decay via the weak interaction before they can hadronize to produce mesons or baryons.
However, the Lorentz-violation scenario in which $\gamma\rightarrow t+\bar{t}$ is possible in vacuum
corresponds to one in which the $t$ and $\bar{t}$ energies grow more slowly with energy than the energy of
the parent photon---that is, $\delta_{t},\delta_{\bar{t}}<\delta_{\gamma}$. A significantly negative
$\delta_{t}$ also corresponds, in fact, to precisely the kind of scenario in which the weak decay of a $t$
might be energetically forbidden. Due to the slower than normal growth of $E_{t}(\vec{p}\,)$, a sufficiently-fast-moving
$t$ might not possess enough energy to be able to produce a virtual $W^{+}$ and a light quark with the same total
momentum!

Thanks to ultrarelativistic beaming, all the constituents in a putative photon decay process like
$\gamma\rightarrow t+\bar{t}$ are moving essentially collinearly; angular deviations near threshold
are typically at a
${\cal O}(m_{t}/E_{\gamma})$ level. This has important implications for the kinematics. Obviously, only the
$\delta(\hat{p})$ parameters for the unique direction of motion will enter into the kinematics. However, there
can also be other, more subtle, effects.

In the decay of a photon into leptons, such as $\gamma\rightarrow e^{+}+e^{-}$, the collinearity limits
which SME coefficients may have observable effects. In this process, without strong final-state interactions,
the exiting quanta must essentially both have the same helicity. Since the initial photon is in a spin-1 state
with allowed helicities $\pm 1$, the two spin-$\frac{1}{2}$ daughter particles must have identical helicities at
threshold, where all the motion is along a single direction.
(Angular-momentum nonconservation is, in principle, possible in Lorentz-violating processes; however, that
would entail an invariant matrix element squared that would be of second order in the Lorentz violation, and
thus negligibly small.)
This means that the $d^{\mu\nu}$ coefficients for the leptons have no effect on the threshold, since any
$d^{\mu\nu}$ term will
shift the dispersion relations for the particle and antiparticle in opposite directions~\cite{ref-altschul17}.
Thus the observation of
the absence of a $\gamma\rightarrow e^{+}+e^{-}$ threshold provides a relatively clean constraint on a linear
combination of just $c_{e}^{\mu\nu}-\frac{1}{2}\tilde{k}_{F}^{\mu\nu}$ components.

The situation is potentially
rather different for a putative photon decay into quarks, because the quarks do not themselves
exist as external states. The ultimate daughter state following $\gamma\rightarrow t+\bar{t}$
will consist of hadronized particles, most generally forming
jets. However, at the threshold, the minimal final-state configuration will include two top-flavored mesons.
The identities of the other quarks present in these mesons are relatively unimportant, since the mesons' masses (and thus
their kinematics) will be totally dominated by the current mass of the top field, $m_{t}\approx 173$ GeV.
However, the mere presence of these
additional quarks, making the daughter particles at threshold composite states, with
more than just the $t$ and $\bar{t}$ carrying angular momentum, changes the analysis so that the $d_{t}^{\mu\nu}$
components
could conceivably also be involved. Unlike a freely propagating lepton, a quark in a bound state does not need to
have a single, well-defined angular momentum state, because the other partons can also carry angular momentum. It
thus may not be instantly obvious that we could not produce $t$-$\bar{t}$ pairs
that will settle down with opposite, rather than aligned,
helicities. In fact, after hadronization, the $t$ and $t$-bar constituents of two different hadrons could have
instantaneously different helicities. However, the exchange of virtual gluons, which are components of a spin-1 field,
between the quark constituents of a hadron ensures that a given quark cannot be maintained in a single consistent
spin state. Inside a spin-0 meson, for instance, the spin of each of the valance quarks must average to zero, and
because of this, the spin-dependent effects of any quark $d^{\mu\nu}$ coefficients must also vanish. We therefore introduce
the spin-averaged quantity
\begin{equation}
\label{eq-delta-bar}
\overline{\delta}(\hat{v})=-c_{00}-c_{(0j)}\hat{v}_{j}-c_{jk}\hat{v}_{j}\hat{v}_{k},
\end{equation}
as it turns out that for quarks, just like leptons, this is
the particular combination that photon survival observations
are sensitive to.

We now must look in detail at the kinematics that might allow for the existence of top-flavored meson daughter
particles in a photon decay.
We consider a configuration with two mesons---one of them $T$, which may have the quark content $t\bar{q}$ for
any other flavor $q$, and the other the antiparticle $\overline{T}$. To the order of magnitude precision we need,
$m_{T}\approx m_{t}$. For the $T$-$\overline{T}$ pair to be produced, the energy of a photon moving in the
$\hat{p}$-direction must exceed
\begin{equation}
\label{eq-ET}
E_{{\rm th}}=\frac{2m_{T}}{\sqrt{2\delta_{\gamma}(\hat{p})-\delta_{T}(\hat{p})-\delta_{\overline{T}}(\hat{p})}}.
\end{equation}
As expected, a real value for the threshold energy $E_{{\rm th}}$ only exists if the MAV parameters satisfy
$2\delta_{\gamma}(\hat{p})-\delta_{T}(\hat{p})-\delta_{\overline{T}}(\hat{p})>0$, indicating a steeper dispersion
relation $E(\vec{p}\,)$ for the photon than for the mesons.

In order that the $c_{t}^{\mu\nu}$ coefficients should determine the kinematics, it must also be the case that the
usual weak decay of the $t$ be forbidden.
The most accessible decay modes of the bare, unhadronized $t$
are of the form $t\rightarrow q+l+\nu$, with a quark, a lepton, and
a neutrino. The lightest decay products occur via a radiative decay,
in which $q=u$ and the lepton is actually a second
neutrino. This is rather unlike the usual $t$ decay modes, which proceed through $t\rightarrow W^{+}+b$, with subsequent
$W^{+}\rightarrow l^{+}+\nu$ or $W^{+}\rightarrow{\rm jets}$. However, under the assumption that the Lorentz
violation in the entire process is dominated by the $c_{t}^{\mu\nu}$ coefficients, the
specific identities of the $t$-decay daughter particles are relatively unimportant, provided they are all much lighter than
$m_{t}$. For the specific $t\rightarrow u+\nu_{1}+\nu_{2}$ decay process to occur, we must have
\begin{equation}
(1+\delta_{t})p_{t}+\frac{m_{t}^{2}}{2p_{t}}>(1+\delta_{u})p_{u}+\frac{m_{u}^{2}}{2p_{u}}+
(1+\delta_{\nu_{1}})p_{\nu_{1}}+(1+\delta_{\nu_{2}})p_{\nu_{2}},
\end{equation}
with $p_{u}+p_{\nu_{1}}+p_{\nu_{2}}=p_{t}$ and
taking the neutrinos to be effectively massless. Presuming a negative $\delta_{t}$ and negligible Lorentz violation
for the neutrinos (flavor-diagonal Lorentz violation in the neutrino sector being well constrained, typically at the
$10^{-17}$ level, by cosmic ray observations~\cite{ref-altschul18,ref-diaz1}),
the threshold configuration
at which equality holds puts virtually all the momentum in the $u$. That means that the $t$ decay is allowed if
\begin{equation}
\label{eq-t-decay}
\delta_{t}-\delta_{u}>\frac{m_{u}^{2}-m_{t}^{2}}{2p_{t}^{2}}\approx-\frac{m_{t}^{2}}{2p_{t}^{2}}.
\end{equation}

Since (\ref{eq-t-decay}) is the condition for the $t$ to decay weakly (via one particular radiative channel)
prior to hadronization, it can be generalized and turned around
to give a stability condition, for a $t$-containing meson $T$ to have time to exist. Direct decay of a $t$  with
momentum $p_{t}$ will be forbidden (and thus it will hadronize) if
\begin{equation}
\label{eq-hadronize}
\overline{\delta}_{t}-\overline{\delta}_{q}<-\frac{m_{t}^{2}}{2p_{t}^{2}},
\end{equation}
where $\delta_{q}$ is the modification to the MAV for any lighter quark that the $t$ might decay into. In fact, the
$c^{\mu\nu}$ coefficients for all the lighter quarks have been previously constrained in various ways. For the $u$ and $d$,
many strong constraints are available, and even for the $s$, $c$, and $b$, there are the observed absences of the
photon decays $\gamma\rightarrow K^{+}+K^{-}$, $\gamma\rightarrow D^{+}+D^{-}$, and $\gamma\rightarrow B^{+}+B^{-}$.
Since all these mesons are much lighter than $m_{t}$, the nonoccurrence of these decays leads to bounds on the
$c_{q}^{\mu\nu}-\frac{1}{2}\tilde{k}^{\mu\mu}$ parameters that are---according to the characteristic
${\cal O}(m_{X}^{2}/E_{\gamma}^{2})$ strength
of the bounds---orders of magnitude better than the $m_{t}$ bounds we are primarily interested in here. Moreover,
the observation of a given source, with source-to-Earth direction $\hat{p}$, will constrain the same linear combinations
of coefficients $\overline{\delta}$ for all the quarks. By the same token, the gluon configuration in a light-heavy meson
such as $D$ or $B$ has been found not to contribute significant Lorentz violation, and thus, to level of precision
relevant here, the stability of the $t$ against immediate decay is indicative of a bounds on just the
combination linear combination $\overline{\delta}_{t}-\delta_{\gamma}$ of $t$ and photon SME parameters.

We are now in position to assemble all our results into final bounds.
Neglecting the electromagnetic Lorentz violation (because of the strong bounds on the $\tilde{k}_{F}$ coefficients),
if the $T$ is a viable asymptotic state, but photons with energy $E_{\gamma}$ moving in the direction $\hat{p}$ do
not decay via $\gamma\rightarrow T+\overline{T}$, then according to (\ref{eq-ET}),
\begin{equation}
\label{eq-deltaT-barT}
\delta_{T}(\hat{p})+\delta_{\overline{T}}(\hat{p})>-\frac{4m_{T}^{2}}{E_{\gamma}^{2}}.
\end{equation}
In an ordinary meson, the valance quarks each carry about one quarter of the momentum, with the other half
belonging to the gluon field. It is straightforward to see from (\ref{eq-v}) that with the two quarks
carrying essentially equal momentum fractions, $x_{t}=p_{t}/p_{T}\approx x_{\bar{q}}=p_{\bar{q}}/p_{T}$, and with
Lorentz violation existing at the threshold level, the
two valance quark constituents are moving at the same velocity---which is the overall velocity of the meson.
So the internal momentum distribution in the $T$ meson is not greatly affected by the Lorentz violation,
at the $T$-$\overline{T}$ decay threshold.
This then simplifies (\ref{eq-deltaT-barT}) to
\begin{equation}
\label{eq-final}
\overline{\delta}_{t}(\hat{p})\gtrsim-\frac{8m_{t}^{2}}{E_{\gamma}^{2}}.
\end{equation}
However, in order for the top-flavored mesons to exists as asymptotic states, the bare $t$ that is initially produced
(with momentum $\frac{1}{2}E_{\gamma}$ at threshold) must not be susceptible to weak decay. According to
(\ref{eq-hadronize}), the condition for the bare $t$ to be stabilized against weak decay, so
that $\gamma\rightarrow T+\overline{T}$ may occur, is actually less stringent than (\ref{eq-final}) by a factor of 4. Thus
(\ref{eq-final}) represents our final analytical result.

\section{Conclusions}
\label{sec-concl}

The form of the bound (\ref{eq-final}) is similar to that for other $\gamma\rightarrow X^{+}+X^{-}$
reactions given in~\cite{ref-altschul14}. The scale of the bounds is comparable, although the more complicated
details associated with the hadronization around the $t$ appear to lead to about a factor of 4 loss in sensitivity.
However, that factor really just corresponds to the difference between the coefficients for the $T$ mesons and
the underlying $t$ quarks (representing the inverse of the momentum fraction carried by the $t$).
Essentially the same factor would exist in comparing the sensitivities of the
$\delta_{B}$ for $B^{\pm}$ mesons, versus the the $\overline{\delta}_{b}$ for their heavy quark.
Moreover, this analysis
still provides bounds that are better than any others available for most of the $c_{t}^{\mu\nu}$ coefficients.
A sample of astrophysical sources of high-energy photons, coming from a wide range of directions, was listed
in~\cite{ref-altschul14}; the energies involved extend up to the tens of TeV. At this scale, the sensitivity scale
is given by $\sim 8m_{t}^{2}/(20\,{\rm TeV})\approx 6\times10^{-4}$.

The one unfortunate feature of these bounds (which a common issue with any constraints based solely on the absence of
the photon decay process), is that all the bounds on the $\bar{\delta}(\hat{p})$ are one sided. By making use of
a wide variety of sources, it is possible to constrain the allowed parameter space significantly more strongly, but it is
evident, for example from (\ref{eq-delta-bar}) and (\ref{eq-final}), that it will
never be possible to rule out a negative value of $c_{t}^{00}$. A Lorentz-violating theory will just a negative
$c_{t}^{00}$ indicates an isotropic speedup of the $t$ excitations, with a steeper dispersion relation for any
top-flavored
quanta than for photons. It is obvious in this case that the anomalous photon decay process will never occur; it
is even more strongly forbidden than in the Lorentz-invariant theory. So the absence of photon decay provides no
information about this corner of the parameter space.

The bound from~\cite{ref-satunin} based on radiative corrections can be seen as complementary to
the collection of photon survival bounds derived from observations of different sources. In particular,~\cite{ref-satunin}
provides a two-sided bound on $c_{t}^{00}$, which is the one specific $c_{t}^{\mu\nu}$ coefficient that cannot,
in principle, be constrained on both sides by the method discussed in this paper. However, those
results and these cannot be directly combined, since the analysis in~\cite{ref-satunin} assumed rotational
isotropy from the start.

The precision of the photon survival bounds will naturally be improved by the observation of ever higher-energy
cosmic ray photons, coming from a wide range of approach directions. However, improvements of this type are
ultimately limited by the availability of such photons, and there is thus an eventual physical cutoff beyond which
the photon survival bounds cannot be further strengthened. For this reason,
the bounds coming from radiative corrections and affecting
photon arrival times may, in the long run, provide a systematically better approach. However, the calculations in
of~\cite{ref-satunin} would need to be extended to apply to situations without spherical isotropy in order for
this improvement to be effected, and there would still be the necessity of observing photon arrival times along
multiple different directions, not merely from a single source.

In this paper, we have generalized the previous analysis of how the observed absence of the photon decay process
$\gamma\rightarrow X^{+}+X^{-}$ can be used to place constraints on Lorentz violation. The generalized analysis has
allowed us to treat possible decays into $t$ quarks, which were not previously included because the $t$
ordinarily decays too quickly to hadronize into a suitable external color-confined state. However, the same
Lorentz-violating kinematics that could make $\gamma\rightarrow t+\bar{t}$ possible would also tend to make the
ordinary weak decay of the bare top energetically impossible. This means that much of the analysis applied to
ordinarily stable daughter particles can be carried over, and the results are that there are bounds on
$t$ sector Lorentz violation coefficients $c_{t}^{\mu\nu}$ at a characteristically $\sim 10^{-4}$ level of
precision.


\begin{thebibliography}{99}

\bibitem{ref-kost1}D. Colladay, V. A. Kosteleck\'{y}, Phys. Rev. D {\bf 55},
6760 (1997).
\bibitem{ref-kost2}D. Colladay, V. A. Kosteleck\'{y}, Phys. Rev. D {\bf 58},
116002 (1998).
\bibitem{ref-tables}V. A. Kosteleck\'{y}, N. Russell, Rev. Mod. Phys. {\bf 83},
11 (2011); updated as arXiv:0801.0287v13.
\bibitem{ref-jacobson1}T. Jacobson, S. Liberati, D. Mattingly, Nature {\bf 424},
1019 (2003).
\bibitem{ref-altschul6}B. Altschul, Phys. Rev. Lett. {\bf 96}, 201101 (2006).
\bibitem{ref-altschul15}B. Altschul, Phys. Rev. D {\bf 75}, 041301 (R) (2007).
\bibitem{ref-altschul16}B. Altschul, Phys. Rev. D  {\bf 77}, 105018 (2008).
\bibitem{ref-schwinger1}J. Schwinger, Phys. Rev. {\bf 82}, 664 (1951).
\bibitem{ref-kost5}V. A. Kosteleck\'{y}, A. G. M. Pickering, Phys. Rev.
Lett. {\bf 91}, 031801 (2003).
\bibitem{ref-mewes5}V. A. Kosteleck\'{y}, M. Mewes, Phys. Rev. Lett. {\bf 99}, 011601 (2007).
\bibitem{ref-mewes6}V. A. Kosteleck\'{y}, M. Mewes, Phys. Rev. Lett. {\bf 110}, 201601 (2013).
\bibitem{ref-mewes7}V. A. Kosteleck\'{y}, A. C. Melissinos, M. Mewes, Phys. Lett. B {\bf 761}, 1 (2016).
\bibitem{ref-altschul4}B. Altschul, D. Colladay, Phys. Rev. D {\bf 71}, 125015
(2005).
\bibitem{ref-hohensee1}M. A. Hohensee, R. Lehnert, D. F. Phillips, R. L. Walsworth,
Phys. Rev. Lett. {\bf 102}, 170402 (2009).
\bibitem{ref-stecker}F. W.  Stecker, S. L.  Glashow, Astropart. Phys. {\bf 16},  97
(2001).
\bibitem{ref-altschul14}B. Altschul, Astropart. Phys. {\bf 28}, 380 (2007).
\bibitem{ref-abazov}V. M. Abazov, {\em et al.} (D0 Collaboration), Phys. Rev. Lett.
{\bf 108}, 261603 (2012).
\bibitem{ref-berger4}M. S. Berger, V. A. Kosteleck\'{y}, Z. Liu, Phys. Rev. D {\bf 93}, 036005 (2016).
\bibitem{ref-kost3}V. A. Kosteleck\'{y}, C. D. Lane, A. G. M. Pickering, 
Phys. Rev. D {\bf 65}, 056006 (2002).
\bibitem{ref-satunin}P. Satunin, Phys. Rev. D {\bf 97}, 125016 (2018).
\bibitem{ref-altschul17}B. Altschul, Phys. Rev. D {\bf 83}, 056012 (2011).
\bibitem{ref-altschul18}B. Altschul, J. Phys. Conf. Ser. {\bf 173}, 012003 (2009).
\bibitem{ref-diaz1}J. S. Diaz, V. A. Kosteleck\'{y}, M. Mewes, Phys. Rev. D {\bf 89}, 043005 (2014).


\end{thebibliography}
\end{document}